\documentclass[final,5p,times,twocolumn]{elsarticle}
\pdfoutput=1

\biboptions{sort&compress}
\journal{Physics Letters B}

\usepackage{aas_macros}

\usepackage[utf8]{inputenc}
\usepackage{graphicx}
\graphicspath{{Plots/}}
\usepackage{amsmath}
\usepackage{amssymb}
\usepackage{bm}
\usepackage{paralist,array}
\usepackage{units} 
\usepackage{xcolor}
\usepackage{slashed}
\usepackage[colorlinks=true,citecolor=blue]{hyperref}

\newcommand{\eV}{{\rm eV}}
\newcommand{\Gauss}{{\rm G}}

\newcommand{\Mpc}{{\rm Mpc}}
\newcommand{\seconds}{{\rm s}}

\newcommand{\rev}[1]{\textcolor{black}{#1}}
\newcommand{\revplb}[1]{\textcolor{red}{#1}}

\colorlet{red}{black}

\begin{document}

\begin{frontmatter}

\title{Neutron stars as photon double-lenses: constraining resonant conversion into ALPs}

\author[l1,l2,l3]{Kyrylo Bondarenko}
 \ead{kyrylo.bondarenko@sissa.it}
\address[l1]{
IFPU, Institute for Fundamental Physics of the Universe, via Beirut 2, I-34014 Trieste, Italy}
\address[l2]{
SISSA, via Bonomea 265, I-34132 Trieste, Italy
}
\address[l3]{
INFN, Sezione di Trieste, SISSA, Via Bonomea 265, 34136, Trieste, Italy
}

\author[l4]{Alexey Boyarsky}
 \ead{boyarsky@lorentz.leidenuniv.nl}
\address[l4]{%
 Institute Lorentz, Leiden University, Niels Bohrweg 2, Leiden, NL-2333 CA, the Netherlands
}

\author[l5,l6,l7]{Josef Pradler}
\ead{josef.pradler@univie.ac.at}
\address[l5]{Institute of High Energy Physics, Austrian Academy of Sciences, Georg-Coch-Platz 2, 1010 Vienna, Austria}
\address[l6]{University of Vienna, Faculty of Physics, Boltzmanngasse 5, A-1090 Vienna, Austria}
\address[l7]{CERN, Theoretical Physics Department, 1211 Geneva 23, Switzerland}

\author[l8,l9]{Anastasia Sokolenko\corref{cor1}}
 \cortext[cor1]{sokolenko@kicp.uchicago.edu}
\affiliation[l8]{
Theoretical Astrophysics Department, Fermi National Accelerator Laboratory, Batavia, Illinois, 60510, USA
}
\affiliation[l9]{
Kavli Institute for Cosmological Physics, The University of Chicago, Chicago, IL 60637, USA
}

\begin{abstract}
Axion-photon conversion is a prime mechanism to detect axion-like particles that share a coupling to the photon. We point out that in the vicinity of neutron stars with strong magnetic fields, magnetars, the effective photon mass receives comparable but opposite contributions from free electrons and the radiation field. This leads to an energy-dependent resonance condition for conversion that can be met for arbitrary light axions and leveraged when using systems with detected radio component. Using the magnetar SGR J1745-2900 as an exemplary source, we demonstrate that sensitivity to $|g_{a\gamma}| \sim 10^{-12}\,\rm{GeV^{-1}}$ or better  can be gained for $m_a \lesssim 10^{-6}\,\rm eV$, with the potential to  improve current constraints on the axion-photon coupling by more than one order of magnitude over a broad mass range. With growing insights into the physical conditions of magnetospheres of magnetars, the method hosts the potential to become a serious competitor to future experiments such as ALPS-II and IAXO in the search for axion-like particles.
\end{abstract}

\begin{keyword}
axion-like particle \sep axion-photon resonant conversion \sep magnetar
\end{keyword}

\end{frontmatter}


\paragraph{Introduction} 
Axions, originally introduced as a solution to the strong CP-problem of QCD~\cite{Peccei:1977hh,Peccei:1977ur,Weinberg:1977ma,Wilczek:1977pj,Kim:1979if,Shifman:1979if,Zhitnitsky:1980tq,Dine:1981rt}, with the additional benefit of serving as dark matter (DM) candidates~\cite{Preskill:1982cy,Abbott:1982af,Dine:1982ah} are now broader appreciated as low-energy messengers of high-scale new physics. \emph{For example}, in string constructions, axion-like particles (ALPs) appear in multitude~\cite{Svrcek:2006yi,Arvanitaki:2009fg,Acharya:2010zx,Ringwald:2012cu,Kamionkowski:2014zda,Stott:2017hvl,Halverson:2019cmy}, populating a great range of masses $m_a$ and couplings, representing viable targets for astronomical and laboratory-based searches~\cite{Raffelt:2006cw,Jaeckel:2010ni,Graham:2015ouw,Marsh:2015xka,Irastorza:2018dyq}. Cumulatively, the largest efforts have gone into probing the axion-photon coupling. Here, once the axion mass is below the $\mu$eV scale, the leading probes are from high energy astrophysics: axions can be produced in stars or supernovae~\cite{Raffelt:1990yz,Giannotti:2017hny} and subsequently convert into X- and $\gamma$-rays in  galactic magnetic fields~\cite{Dessert:2020lil,Xiao:2020pra,Payez:2014xsa}. Alternatively, photon-ALP oscillations can leave their imprint on high energy photon spectra, e.g.~leading to the dimming of sources~\cite{Mirizzi:2007hr,Hooper:2007bq,Hochmuth:2007hk,DeAngelis:2007wiw}, or, sometimes, even explain their brightness~\cite{DeAngelis:2007dqd,Horns:2012kw}.

Some of the strongest constraints limiting the axion-photon coupling to $|g_{a\gamma}|\lesssim 10^{-12}\,{\rm GeV^{-1}}$ have been placed from  \emph{non-resonant} conversion using well-measured spectra of radio galaxies, such as from NGC~1275 in the Perseus cluster \cite{Berg:2016ese,Reynolds:2019uqt} and  M87~\cite{Marsh:2017yvc} in the Virgo cluster. 
In turn, {\it resonant} conversion has been used to search for axion DM through radio lines produced in the conversion to photons using the strong magnetic fields of neutron stars (NS)~\cite{Hook:2018iia,Pshirkov:2007st,Huang:2018lxq,Safdi:2018oeu,Foster:2020pgt,Darling:2020plz,Battye:2021xvt}, and significant recent effort is being invested in pinning down the physical circumstances of signal formation~\cite{Witte:2021arp,Battye:2021yue,Millar:2021gzs}. 
Of central importance to these studies is the account of the medium-dependent mixing between photons and axions~\cite{Raffelt:1987im}. Photons experience a modification of their dispersion relation in media that may be cast in terms of an effective mass $m_{\text{eff}}$. As is well known, the forward scattering on free, non-relativistic charges leads to a positive, energy-independent contribution $m_{\text{eff}}^2|_{\text{charge}} \simeq \omega_p^2$, where  $\omega_p$ is the plasma frequency~\cite{Braaten:1993jw}. Resonant conversion becomes possible when $m_a^2 = m_{\rm eff}^2$, and is hence believed to be relevant only in a narrow range of ALP mass where $m_a^2=\omega_p^2$ is met along the propagation path associated with significant conversion probability.

The purpose of this paper to point out that \emph{resonant} conversion  is possible over a large range in $m_a$, considerably improving the sensitivity to~$g_{a\gamma}$ and independent of the condition that ALPs constitute DM. The central observation is that photons also receive an energy-dependent, negative contribution to their dispersion relation from the background radiation field~\cite{Dobrynina:2014qba},  $m_{\rm eff}^2|_{\rm EM}(\omega) <0 $, where $\omega$ is the photon energy. This has important ramifications for photon-ALP conversion: because the contribution grows in magnitude with photon energy as $\omega^2$, the resonance condition $m_a^2 = m_{\rm eff}^2(\omega)$ that can eventually be met for \emph{any} axion mass $m_a\leq  \omega_p $. 
In the magnetospheres of NS with the strongest magnetic fields, magnetars, this condition is met for radio frequencies (GHz-THz) close to the surface. As the strength of resonances increases with energy, the flux of photons can reduce significantly. Together with the presence of a geometric boundary given by the NS surface, this imprints a sharp spectral feature that can be searched for in radio data. Our objective here is to point out the main ideas and study the principal sensitivity to $g_{a\gamma}$. It is to be followed up by more detailed analyses of the involved stellar systems~\cite{companion}.

The paper is organized as follows. First, we establish the resonance conditions and the expected photon flux. We then demonstrate our ideas using the radio measurements of the magnetar SGR J1745-2900 close to the Galactic center before concluding.

\begin{figure}
    \centering
    \includegraphics[width=\linewidth]{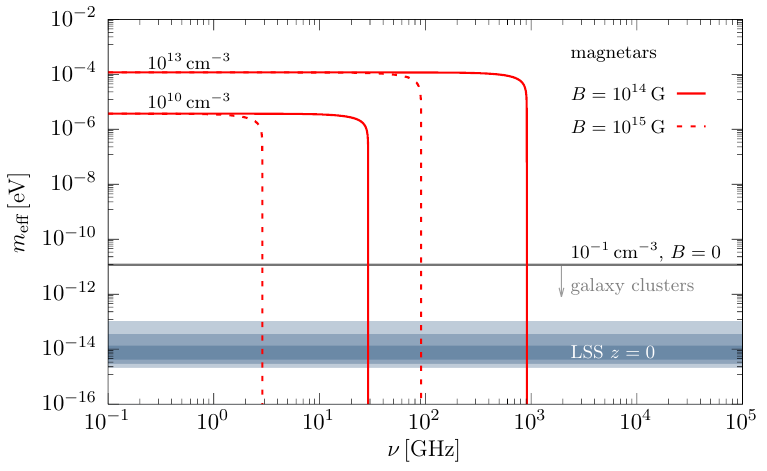}
    \caption{Real branch of the effective photon mass  as a function of photon frequency $\nu$ and  electron densities as labeled. The solid (dashed) red lines represent a neutron star environment with $B=10^{14}$~G ($10^{15}$~G). For comparison, the gray horizontal line shows a typical galaxy cluster central density~\cite{Sarazin:1986zz}. The  1$\sigma$-3$\sigma$ shaded bands are  inferred from the free electron distribution of the large scale structure at $z=0$~\cite{Garcia:2020qrp}.}
    \label{fig:meff2}
\end{figure}

\paragraph{Resonant axion conversion}
Axion-like particles $a$ interact with photons through the effective Lagrangian,
\begin{equation}
\label{Lint}
    \mathcal{L}_{\rm int} = - \frac{g_{a\gamma}}{4} a F_{\mu\nu} \tilde{F}^{\mu\nu} = g_{a\gamma} a \vec E \cdot \vec B,
\end{equation}
where $ F_{\mu\nu}$ ($ \tilde{F}^{\mu\nu} $) is the (dual) photon field strength; $g_{a\gamma}$ and $m_a$ are the only other relevant model parameters. For the QCD axion $g_{a\gamma} \sim \alpha/(\pi f_a)$ and  $m_a \approx m_\pi f_\pi/f_{a}$ holds, whereas for ALPs one keeps $m_a$ as a free parameter while $g_{a\gamma}^{-1}$ is still expected to be informative on the UV scale $f_a$ of symmetry breaking; we use the term axion and ALP interchangeably. 
In a magnetic field $\vec B$ the interaction~\eqref{Lint} enables the conversion of photons (with electric field $\vec E$) into axions and vice verse~\cite{Sikivie:1983ip,Sikivie:1985yu,Raffelt:1987im}. If the condition $m_a^2=m_{\text{eff}}^2$ is satisfied along the photon's path, such conversion becomes resonant. 

\rev{The energy dependent photon-axion conversion probability at the point of resonance depends on the orientation of the (anisotropic) magnetic field. Furthermore, a resonance could be met several times, and in a highly magnetized, potentially semi-relativistic  plasma the propagating photon eigenmodes differ from those of a non-relativistic unmagnetized one. The accurate calculation of the photon-axion conversion can hence be a formidable theoretical problem~\cite{Millar:2021gzs}. However, once the conversion probability becomes $O(1)$, those details tend to fade. For example, if several strong resonances were met, 
%
 photon and axion populations are driven towards equipartition, and as is generally true about any  equilibrium state, information about the history how it got there is lost. Similarly, if a single, strong adiabatic conversion were to damp out the entire flux, the only information retained is encoded in the frequency range of where this happens.
 As we shall see below, we indeed expect to enter such saturated regime by virtue of an increasing conversion probability with frequency. Here, the \emph{smallest} saturated conversion with $1/3$ probability is attained when both photon polarization degrees were to take part with comparable strength.
 We therefore employ the simplified conservative model for
the overall photon-to-axion conversion probability~\cite{Grossman:2002by}}
\begin{equation}
P_{\text{tot}} \approx 
    \frac{1}{3}\left(1 - e^{-3P_{\text{lin}}}\right),~~~
     P_{\text{lin}} = \frac{\pi g_{a\gamma}^2 \omega}{m_{a}^2} \sum_i B_{T,i}^2 R_{i}.
    \label{eq:ptot}
\end{equation}
Here, $P_{\rm lin}$ is found from the Landau-Zener transition probability, $B_{T,i}=B_T(\ell_i)$ is the component of the magnetic field orthogonal to the direction of photon/axion propagation and
$R_i = \left|d \ln m_{\text{eff}}^2/d\ell \right|^{-1}_{\ell = \ell_{i}}$
 is the scale parameter
with $\ell$ being the distance along the line-of-sight; negligible redshift has been assumed. The derivative of $R$  is calculated at each point of  resonance  $ m_a^2 = m_{\text{eff}}^2(\ell_{i})$.
The loss of photons into axions is imprinted onto an initial photon flux $F_{\text{in}}(\omega)$ as,
\begin{equation}
\label{modspec}
    F_{\text{obs}}(\omega) = F_{\text{in}}(\omega) [1 - P_{\text{tot}}(\omega)].
\end{equation}
For unpolarized sources, the effective photon mass  
receives two important contributions (``double lens''),
\begin{equation}
   m^2_{\text{eff}} = \omega_p^2 + m^2_{\rm eff}|_{\rm EM}, \quad  m^2_{\rm eff}|_{\rm EM} = - \frac{88\alpha^2 \omega^2}{135 m_e^4} \rho_{\rm EM},
    \label{eq:meff}
\end{equation}
where $\omega_p^2 = 4\pi \alpha n_e/m_e$.
The second, negative term is the photon-photon scattering contribution $m_{\rm eff}^2|_{\rm EM}$ of the radiation field~\cite{Dobrynina:2014qba}. It implies that the resonance condition $m_a^2=m_{\rm eff}^2$ can eventually be met for \emph{any} axion mass satisfying $m_a\leq \omega_p$. This observation is instrumental to our proposal. 
\textcolor{red}{In a supplement, we show that while the condition of the resonant conversion in the strongly magnetized plasma is modified (see e.g.~\cite{Millar:2021gzs}), Eq.~\eqref{eq:meff} and condition $m_a^2 = m_{\rm eff}^2$ serves as a good representation of the resonance condition over a wide range of relative angles of magnetic field and photon propagation. It allows us to exhibit the central ideas without entering the difficult subject of detailed geometric dependencies; for a quantitative discussion of  the latter we refer the reader to the supplementary material.}

Before proceeding, we emphasize that the applicability of~\eqref{eq:ptot} and hence the formulation of resonant conversion requires a well-separated two-level system away from the resonance point, i.e., an off-diagonal mixing that is much smaller than its diagonals in the Hamiltonian. This translates into the condition
$    \epsilon \equiv {2 |g_{a\gamma}| B_{T,i} \omega}/{| m^2_{\text{eff}}(\ell) - m_a^2|} \ll 1 $ for $ |\ell -\ell_i|\gg 0$
and it guides the selection of potential sources and frequency bands for observing axion-photon conversion. If we are---for the sake of the argument---to neglect $m_{\rm eff}^2$ in~$\epsilon$, we may write for a single resonance $P_{\rm lin} \sim \epsilon |g_{a\gamma}| B R$. 
This shows that to compensate for $\epsilon \ll 1$ one should maximize the product $B R$. 
Saturating $R$ by the size $L$ of the system, we estimate for the product $B L\sim 10^{-8}\,\Gauss\,\Mpc$ for the Milky Way and  $ 10^{-6}\,\Gauss\,\Mpc$ for galaxy clusters. %
In the centers of clusters, the contribution from the photon-photon scattering in~\eqref{eq:meff} is of the same size as the plasma frequency $\omega_p^2 \sim m_{\rm eff}^2|_{\text{EM}}$ for GeV energies but the condition $\epsilon \ll 1$ is violated. If we look at magnetars with their extremely large magnetic field strengths $B\sim 10^{14}\,\Gauss$ and size $L\sim 10$~km, we find  $ B L\sim 10^{-4}\,\Gauss\,\Mpc$ while retaining $\epsilon\ll 1$ for GHz-THz frequencies.%
\footnote{\rev{Resonances of the type~\eqref{eq:meff} with qualitative different phenomenology have also been considered for keV-scale \emph{thermal} radiation emanating from the dense neutron star atmospheres; see e.g.~\cite{Lai:2001di,Lai:2006af} and references therein.}}
This demonstrates that NS have a unique potential to probe the smallest values of $g_{a\gamma}$ through resonant conversion, and for the remainder of the paper we focus on this case.

\paragraph{Magnetosphere model and expected signal}
For an estimate of the electron density we use the Goldreich-Julian (GJ) model~\cite{Goldreich1969} that predicts the required magnetospheric co-rotating spatial charge density~$n_c$ as the difference of positive and negative charge carriers,
\begin{equation}
    n_c = \frac{\vec{\Omega}\cdot \vec{B}}{\sqrt{\pi \alpha}} \frac{1}{1 - \Omega^2 r^2 \sin^2\theta},
    \label{eq:GJ-model}
\end{equation}
where $\vec \Omega $ is the vector along the axis of rotation with magnitude  $\Omega = 2\pi/P$ the NS angular frequency ($P$ is a rotation period);
$ \theta $ is the  angle between $\vec{r}$ and $\vec{\Omega}$. We then take $n_e = |n_c|$ as an estimate on the electron (or positron) number density; we caution that 
\rev{larger values from one~\cite{Sobyanin:2016acr}
to several~\cite{Lyutikov:2007fn,Timokhin:2015dua,Cruz:2020vfm} orders of magnitude are argued for.}%
\footnote{\rev{This may in part be compensated by relativistic corrections in electron-positron plasma, $\omega_p^2 = 4\pi \alpha n_e/m_e \langle \gamma_e^{-3}\rangle $ where $\langle \gamma_e^{-3}\rangle  \sim \langle \gamma_e \rangle^{-1} = 10-10^3$~\cite{1998PhRvE..57.3399G}}} 
On the account that $\Omega r \ll 1$ out to large radii, the electron density only depends on the $\hat z$-component of the magnetic field,  $  n_e \simeq \Omega B |\cos\theta_B| /{\sqrt{\pi \alpha}}$, where $\theta_B$ is the angle between $\vec B$ and~$\vec \Omega$.

We are now in a position to study the  condition $m_{\rm eff}^2 = m_a^2$ from~\eqref{eq:meff}. Using $\rho_{\rm EM}=B^2/2$, we may write it as 
$ m_a^2 = C_1 |\cos\theta_B| B - C_2 \omega^2 B^2$
with $ C_1 = \frac{4\Omega\sqrt{\pi \alpha}}{m_e}$ and $C_2 = \frac{44\alpha^2}{135 m_e^4}.$ Being a quadratic equation in $B$, we observe that 
 resonant conversion is only possible for energies below a critical value
\begin{equation*}
    \omega_c = \frac{C_1 |\cos\theta_B|}{2 m_a\sqrt{C_2}} \approx 10^{-2}\text{ eV} 
    \,|\cos\theta_B| 
    \left(\frac{1\text{ s}}{P}\right)
    \left(\frac{10^{-5}\text{ eV}}{m_a}\right),
\end{equation*}
which also highlights that, in the GJ model,  the charge density vanishes for $\cos\theta_B=0$. 
For $\omega<\omega_c$ there are two physical solutions for the resonant magnetic field value. 
In the limit of small frequencies, the photon-photon scattering contribution can always be neglected and one obtains a resonance that is associated with a small magnetic field value at a distance from the NS surface,
\begin{align*}
    B_- &\approx \frac{m_a^2}{C_1 |\cos\theta_B|} \approx \frac{ 10^{12}\text{ G}}{|\cos\theta_B|} 
    \left(\frac{P}{1\text{ s}}\right)
    \left(\frac{m_a}{10^{-5}\text{ eV}}\right)^2.
    \end{align*}
In turn, for growing photon energy, the cancellation between plasma frequency and the photon-photon scattering term becomes possible (``double lens'').
In the limit $\omega\ll\omega_c$ any parametric dependence on $m_a$ can be neglected and the solution is given by,
    \begin{align*}
    B_+ &\approx \frac{C_1 |\cos\theta_B|}{C_2 \omega^2} \approx  10^{15}\text{ G}\, |\cos\theta_B| \left(\frac{1\text{ s}}{P}\right) \left(\frac{10^{-3}\text{ eV}}{\omega}\right)^2.
\end{align*}
 For $\omega\gtrsim10^{-3}$~eV, the required $B_+$ field values are found at radii close to the magnetar's surface.
 They are associated with significant efficiency of conversion and $P_{\rm tot}\approx 1/3$ can be attained.
 Hence, a {\it sharp feature at $\omega=\omega_{\rm kink}$  on the emanating photon flux can be imprinted once the $B_+$ resonance is found}. The second sharp feature is at $\omega = \omega_c$ but is  difficult to access observationally; note that $\omega = 10^{-3}\ (10^{-2})\,\eV$ corresponds to $\nu \simeq 240$~GHz (2.4~THz). Finally, we note that for large $g_{a\gamma}$ and $m_a$ values, the $B_-$ resonance may become efficient enough to deplete the photon flux already for $\omega < \omega_{\rm kink}$, with the general effect of washing out the spectral feature at $\omega_{\rm kink}$; we take this into account below.

To make progress, we may follow previous investigations~\cite{Pshirkov:2007st,Huang:2018lxq,Safdi:2018oeu,Foster:2020pgt,Darling:2020plz,Battye:2021xvt} in assuming that the magnetic field is well described by a dipole configuration,
\begin{equation}
\label{Bdipole}
    \vec{B} = \frac{1}{4\pi} \left[ \frac{3\vec{r}(\vec{r}\cdot \vec{m})}{r^5} - \frac{\vec{m}}{r^3} \right],
\end{equation}
with magnetic moment $\vec{m} = 2\pi B_0 r_0^3 \hat{n}$, where $B_0$ is the maximum field value at the surface of magnetic poles,  $r_0=10$~km is the assumed radius of the neutron star, and $\hat{n}$ is the unit vector along $\vec m$. 
Taking the star's rotation axis along the $\hat z$-direction and the magnetic moment misaligned by an angle $\theta_m $ one arrives at 
$\vec{n}(t) = (\sin\theta_m \cos\Omega t, \sin\theta_m \sin\Omega t, \cos\theta_m).$ Plugging this expression into~\eqref{Bdipole} yields the time-dependent magnetic field. 

This model has several features. First,
for lines-of-sight that end on the polar regions%
\footnote{\rev{Ray-tracing~\cite{Hook:2018iia,Leroy:2019ghm,Battye:2021xvt,Witte:2021arp} in our analysis is not required as we are strictly concerned with photon frequencies larger than~$\omega_p$ and photon deflection from a straight line is small.}}
, the conversion probability is suppressed because of the parallel magnetic field structure, $B_T\ll  B_0$. Second, the GJ model has a quadrupolar structure with directions of vanishing charge density. Both these features suggest strong geometric dependencies. However, to a certain degree, they can be considered artifacts. First, there can be 
\rev{small-scale magnetic features of comparable strength close to the NS surface~\cite{Kaspi:2017fwg},}
%
the region most relevant to us. Second, the actual electron density may differ from the GJ one. 
From Fig.~\ref{fig:meff2} one observes that an increase of $n_e=10^{13}\ {\rm cm^{-3}}$ by two orders of magnitude shifts the spectral break by one order of magnitude into the observationally difficult THz regime. Our proposal hence hinges on electron densities that are not too different from the GJ ones paired with magnetic field values in the $10^{14}-10^{15}\ \Gauss$ ballpark, inducing a spectral break at several hundred GHz or lower. 

\rev{The above findings are derived from~\eqref{eq:meff} which is applicable to unpolarized radiation and under the assumption that the radio signal is generated close to the NS surface. More generally, the photon-photon scattering contribution is birefringent~\cite{Tsai:1974fa,Tsai:1975iz}, and in an ultra-strong magnetized plasma, it is the Langmuir-O mode with both longitudinal and transverse components that is being converted into axions. 
%
%
Indeed, recent proposals put the origin of radio emission---produced in form these O-modes---very close to the surface of pulsars~\cite{Philippov:2014mqa},} provided that such mechanism is also operative in the pulsed radio emission from magnetars---a process that is still poorly understood~\cite{Turolla:2015mwa}. \revplb{Ultimately, our proposal relies on a sufficient mixing between axion and emergent photon radiation. Although favorable examples such as the one mentioned above exist, we do not understand the physics of magnetars well enough to make a robust claim in the positive and a caveat on polarization-dependence persists.}
\rev{Nevertheless, in the supplemental material we show that the dependence on the assumption of the location of radio emission is rather mild: when the radio signal is created at $2r_0$ or $3r_0$, instead of very close to the surface, the sensitivity is only mildly affected. }


\revplb{Finally, in recent work~\cite{Carenza:2023nck}, a modification of the Landau-Zener formula due to the finite region of resonant conversion and fluctuation of magnetic field was discussed for the condition when plasma frequency $\omega_{\text{pl}}$ equals axion mass $m_a$. In the current context, the resonant condition $m^2_{\text{eff}} = m_a^2$, is fulfilled when two large contributions--- plasma frequency and vacuum polarization---cancel each other with high precision.
We may estimate the the radial extend of the region where this cancellation happens with the precision of axion mass,
\begin{equation}
    \Delta r = \frac{m_a^2}{\left|\dfrac{d m^2_{\text{eff}}}{d r}\right|} \approx 10\ \text{m}\  \left(\frac{m_a}{10^{-6}\text{ eV}}\right)^2
    \left(\frac{10^{-3}\text{ eV}}{\omega}\right)^2 \left(\frac{10^{14}\text{ G}}{B}\right)^2 .
\end{equation}
We see that for characteristic numbers used in this work, the size of the resonance region is much smaller than the characteristic size of the neutron star ($10$~km) and the assumed zone of conversion (few km). We therefore expect the Landau-Zener formula to be applicable.}

\begin{figure}
    \centering
    \includegraphics[width=\linewidth]{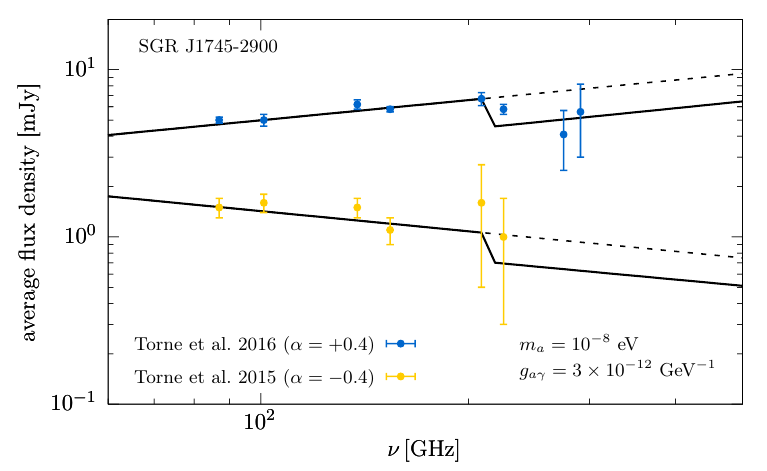}
    \caption{High-frequency radio spectrum of SGR J1745-2900 during two observational campaigns over several days in 2014~\cite{Torne:2015rha} and 2015~\cite{Torne:2016zid} together with reported fitted power laws that include additional data points below 10~GHz (not shown.) The solid lines show the effect of photon-axion conversion for $m_a\leq 10^{-8}$~eV and $g_{a\gamma} = 3\times 10^{-12}\,{\rm GeV}^{-1}$.}
    \label{fig:torne} 
\end{figure}

Having mentioned important caveats to our proposal, we now proceed studying its sensitivity potential
 by making some simplifying assumptions.
 Importantly, in the regime where the conversion becomes saturated, $P_{\rm tot} \approx 1/3$, all geometric dependencies are distilled into the position of the sharp feature in frequency only. In fact, the latter does not depend on direction but only on $B$-field magnitude at the respective region very close to the surface, and we expect $\omega_{\rm kink}$ to remain preserved over the typical observational time windows of several hours.
 Therefore, for the purpose of illustration, we keep the asymptotic radial scaling of a magnetic dipole, $B = B_0(r_0/r)^3$, but take its direction to be random. We replace occurrences of the angle $\theta_B$ by its average assuming its uniform distribution: $ \langle |\cos\theta_B| \rangle = 1/2$ so that $n_e = \Omega B/e$, and take $ \langle \sin^2\theta_B \rangle = 2/3$ in the conversion probability. Taken together, these assumptions allow for a simple exposition of our ideas while retaining the essential features.

\paragraph{SGR J1745-2900 as an exemplary source} 
We choose the radio-loud magnetar SGR J1745-2900, 0.1~pc near the galactic center with a period $P=3.76\,\seconds$~\cite{Kennea:2013dfa,Mori:2013yda} and $B_0=1.6\times 10^{14}\,\Gauss$~\cite{Mori:2013yda} as an exemplary source. Its pulsed radio emissions with mJy flux density have been measured over an unprecedented broad range from 2.54~GHz (118~mm) up to 225~GHz (1.33~mm)~\cite{Torne:2015rha} and to 291~GHz (1.03~mm)~\cite{Torne:2016zid} over a period of several days in 2014 and 2015, respectively. The observed mean spectral densities were relatively flat, with respective power law indices $\langle \alpha \rangle = - 0.4\pm 0.1 $ and  $\langle \alpha \rangle = + 0.4\pm 0.2 $, with a possibility of a spectral break at tens of GHz.
Flux density and spectral index variabilities are observed on long and short time scales \rev{and there is evidence for some degree of linear polarization up to the highest frequencies~\cite{Torne:2016zid}.}
The details of pulsed (radio) emission from magnetars remain poorly understood but are generally expected to be associated with open field lines of polar regions~\cite{Beloborodov:2008df}.%

%

\begin{figure}[t]
    \centering
    \includegraphics[width=\linewidth]{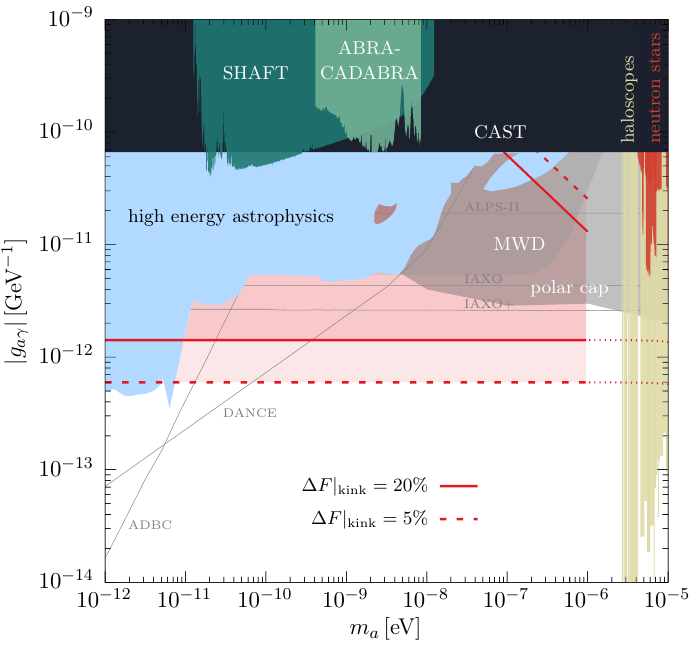}
    \caption{Sensitivity region on photon-ALP resonant conversion bounded by the solid (dashed) red line based on the assumption that a 20\% (5\%) spectral feature can be detected. Astrophysical constraints are cumulatively shown by the blue shaded region labeled ``high energy astrophysics'' (see~\cite{githublimits}) from magnetic white dwarfs (MWD)~\cite{Dessert:2022yqq} \revplb{and pulsar polar caps~\cite{Noordhuis:2022ljw}.} Laboratory limits from CAST~\cite{CAST:2017uph}, SHAFT~\cite{Gramolin:2020ict}, ABRACADABRA~\cite{Salemi:2021gck} and projections for ALPS-II~\cite{Ortiz:2020tgs}, IAXO(+)~\cite{2013ITAS...23T0604S}, DANCE~\cite{Michimura:2019qxr}, and ADBC~\cite{Liu:2018icu} are shown as labeled. Additional constraints that assume ALPs being DM are from haloscopes~\cite{PhysRevLett.59.839,HAYSTAC:2018rwy,ADMX:2019uok,HAYSTAC:2020kwv,Jeong:2020cwz,Lee:2020cfj,ADMX:2021nhd}, previous analyses using neutron stars~\cite{Foster:2020pgt,Darling:2020plz,Battye:2021yue}, \revplb{or axion stars~\cite{Escudero:2023vgv}}.
    \label{fig:plane}}
\end{figure}

Figure~\ref{fig:torne} shows the averaged radio spectral densities of SGR~J1745-2900 as a function of frequency of the two observational campaigns over several days in 2014~\cite{Torne:2015rha} and 2015~\cite{Torne:2016zid}. The dashed lines show the reported fitted power laws (which are additionally anchored by low-frequency data from 1-10~GHz.) The solid line is obtained by multiplying the fits by $1-P_{
\rm tot}$ for $m_a=10^{-8}\,\eV$ and $g_{a\gamma}=3\times 10^{-12}\,{\rm GeV}$ using the simplified magnetospheric model described above. As can be seen, for $\nu \gtrsim 200~{\rm GHz}$, the resonance associated with $B_+$ can be met, leading to a sharp saturated reduction of the flux by a factor one third. 

Figure~\ref{fig:plane} explores the sensitivity in the $(m_a,|g_{a\gamma}|)$ plane to resonant conversion assuming that a sharp spectral feature, i.e., a sudden flux reduction $\Delta F|_{\rm kink}$ can be detected at the frequency where $B_+$ becomes available. 
The solid (dashed) red line depicts $\Delta F|_{\rm kink} = 20\%$ (5\%)\revplb{, which corresponds to the exclusion condition $P_{\text{tot}} > 20\%\ (5\%)$. The former thereby represents a minimal required sensitivity while the latter illustrates the reach of a better instrument.} As can be seen, for $m_a\lesssim 10^{-5}\,\eV$, the result is independent of axion mass and sensitivity as good as $|g_{a\gamma}|\lesssim 1.4\times 10^{-12}\ (6\times 10^{-13})\,{\rm GeV^{-1}}$ can be reached.
For $m_a\gtrsim 10^{-6}\,\eV$ and/or for large $|g_{a\gamma}|$, the conversion associated with $B_-$ is strong enough to deplete the photon flux for frequencies below the onset of $B_+$. This washes out the kink and the red shading indicates the region unaffected by it. 
Existing ALP constraints and other laboratory projections are additionally shown as labeled. As can be seen, an improvement over current limits by more than an order of magnitude is possible, putting astrophysics in competition with upcoming laboratory searches that target a wide range of ALP masses, in particular from $10^{-11}$~eV to $10^{-6}$~eV. 

\paragraph{Conclusions}
 In this work, we show that observations of the high-frequency end of the radio band of magnetars host the possibility to put very stringent constraints on the ALP-photon coupling from resonant conversion at the $10^{-12}\,{\rm GeV^{-1}}$ level. The method works for an arbitrarily small and hence wide range of $m_a$, leveraging the energy-dependent negative contribution from the radiation field to the effective photon mass (``double lens effect''). It removes the direct relation between axion mass and resonant electron number density and, at the same time, places the resonance radius close to the neutron star surface with ensuing strong conversion probability. Because of the sharp NS surface boundary, a spectral feature that can be searched for in high-quality radio data is imprinted.

Currently, an incomplete understanding of the magnetospheres' physical conditions and the ensuing production and propagation of radio photons from there prevents us from claiming real limits. However, many uncertainties are expected to be mitigated by future observations with SKA~\cite{Kramer:2006ha,Watts:2014trg,Keane:2014vja,Karastergiou:2014cka,Antoniadis:2015nga}, complemented by the steady stream of high-frequency observations in the few hundred~GHz range by mm/sub-mm arrays such as ALMA, IRAM, or JCMT~\cite{Chu:2021ktf,Torne:2021yad,Torne:2022xfo}, and through paralleling advances in the simulation and modeling of these extreme objects~\cite{Philippov:2014mqa,Petri:2016tqe,Chen:2016qkk,Cerutti:2016ttn,Brambilla:2017puc,Carrasco:2020sxg,2020ApJ...889...69C}. Together, photon-axion conversion in neutron stars may well become a serious competitor to the experimental ALPS-II and IAXO programs.

\paragraph{Acknowledgements}
We thank Yuri Levin and Andrii Neronov for helpful discussions.
KB is partly funded by the INFN PD51 INDARK grant.
AB is supported by the European Research Council (ERC) Advanced Grant ``NuBSM'' (694896). AS is supported  by the Kavli Institute for Cosmological Physics at the University of Chicago through an endowment from the Kavli Foundation and its founder Fred Kavli. This work has been supported by the Fermi Research Alliance, LLC under Contract No. DE-AC02-07CH11359 with the U.S. Department of Energy, Office of High Energy Physics.

\bibliographystyle{JHEP}
\bibliography{refs.bib}

\providecommand{\href}[2]{#2}\begingroup\raggedright\begin{thebibliography}{100}

\bibitem{Peccei:1977hh}
R.~D. Peccei and H.~R. Quinn, {\it {CP Conservation in the Presence of
  Instantons}},  {\em Phys. Rev. Lett.} {\bf 38} (1977) 1440--1443.

\bibitem{Peccei:1977ur}
R.~D. Peccei and H.~R. Quinn, {\it {Constraints Imposed by CP Conservation in
  the Presence of Instantons}},  {\em Phys. Rev. D} {\bf 16} (1977) 1791--1797.

\bibitem{Weinberg:1977ma}
S.~Weinberg, {\it {A New Light Boson?}},  {\em Phys. Rev. Lett.} {\bf 40}
  (1978) 223--226.

\bibitem{Wilczek:1977pj}
F.~Wilczek, {\it {Problem of Strong $P$ and $T$ Invariance in the Presence of
  Instantons}},  {\em Phys. Rev. Lett.} {\bf 40} (1978) 279--282.

\bibitem{Kim:1979if}
J.~E. Kim, {\it {Weak Interaction Singlet and Strong CP Invariance}},  {\em
  Phys. Rev. Lett.} {\bf 43} (1979) 103.

\bibitem{Shifman:1979if}
M.~A. Shifman, A.~I. Vainshtein, and V.~I. Zakharov, {\it {Can Confinement
  Ensure Natural CP Invariance of Strong Interactions?}},  {\em Nucl. Phys. B}
  {\bf 166} (1980) 493--506.

\bibitem{Zhitnitsky:1980tq}
A.~R. Zhitnitsky, {\it {On Possible Suppression of the Axion Hadron
  Interactions. (In Russian)}},  {\em Sov. J. Nucl. Phys.} {\bf 31} (1980) 260.

\bibitem{Dine:1981rt}
M.~Dine, W.~Fischler, and M.~Srednicki, {\it {A Simple Solution to the Strong
  CP Problem with a Harmless Axion}},  {\em Phys. Lett. B} {\bf 104} (1981)
  199--202.

\bibitem{Preskill:1982cy}
J.~Preskill, M.~B. Wise, and F.~Wilczek, {\it {Cosmology of the Invisible
  Axion}},  {\em Phys. Lett. B} {\bf 120} (1983) 127--132.

\bibitem{Abbott:1982af}
L.~F. Abbott and P.~Sikivie, {\it {A Cosmological Bound on the Invisible
  Axion}},  {\em Phys. Lett. B} {\bf 120} (1983) 133--136.

\bibitem{Dine:1982ah}
M.~Dine and W.~Fischler, {\it {The Not So Harmless Axion}},  {\em Phys. Lett.
  B} {\bf 120} (1983) 137--141.

\bibitem{Svrcek:2006yi}
P.~Svrcek and E.~Witten, {\it {Axions In String Theory}},  {\em JHEP} {\bf 06}
  (2006) 051, [\href{http://arxiv.org/abs/hep-th/0605206}{{\tt
  hep-th/0605206}}].

\bibitem{Arvanitaki:2009fg}
A.~Arvanitaki, S.~Dimopoulos, S.~Dubovsky, N.~Kaloper, and J.~March-Russell,
  {\it {String Axiverse}},  {\em Phys. Rev. D} {\bf 81} (2010) 123530,
  [\href{http://arxiv.org/abs/0905.4720}{{\tt arXiv:0905.4720}}].

\bibitem{Acharya:2010zx}
B.~S. Acharya, K.~Bobkov, and P.~Kumar, {\it {An M Theory Solution to the
  Strong CP Problem and Constraints on the Axiverse}},  {\em JHEP} {\bf 11}
  (2010) 105, [\href{http://arxiv.org/abs/1004.5138}{{\tt arXiv:1004.5138}}].

\bibitem{Ringwald:2012cu}
A.~Ringwald, {\it {Searching for axions and ALPs from string theory}},  {\em J.
  Phys. Conf. Ser.} {\bf 485} (2014) 012013,
  [\href{http://arxiv.org/abs/1209.2299}{{\tt arXiv:1209.2299}}].

\bibitem{Kamionkowski:2014zda}
M.~Kamionkowski, J.~Pradler, and D.~G.~E. Walker, {\it {Dark energy from the
  string axiverse}},  {\em Phys. Rev. Lett.} {\bf 113} (2014), no.~25 251302,
  [\href{http://arxiv.org/abs/1409.0549}{{\tt arXiv:1409.0549}}].

\bibitem{Stott:2017hvl}
M.~J. Stott, D.~J.~E. Marsh, C.~Pongkitivanichkul, L.~C. Price, and B.~S.
  Acharya, {\it {Spectrum of the axion dark sector}},  {\em Phys. Rev. D} {\bf
  96} (2017), no.~8 083510, [\href{http://arxiv.org/abs/1706.03236}{{\tt
  arXiv:1706.03236}}].

\bibitem{Halverson:2019cmy}
J.~Halverson, C.~Long, B.~Nelson, and G.~Salinas, {\it {Towards string theory
  expectations for photon couplings to axionlike particles}},  {\em Phys. Rev.
  D} {\bf 100} (2019), no.~10 106010,
  [\href{http://arxiv.org/abs/1909.05257}{{\tt arXiv:1909.05257}}].

\bibitem{Raffelt:2006cw}
G.~G. Raffelt, {\it {Astrophysical axion bounds}},  {\em Lect. Notes Phys.}
  {\bf 741} (2008) 51--71, [\href{http://arxiv.org/abs/hep-ph/0611350}{{\tt
  hep-ph/0611350}}].

\bibitem{Jaeckel:2010ni}
J.~Jaeckel and A.~Ringwald, {\it {The Low-Energy Frontier of Particle
  Physics}},  {\em Ann. Rev. Nucl. Part. Sci.} {\bf 60} (2010) 405--437,
  [\href{http://arxiv.org/abs/1002.0329}{{\tt arXiv:1002.0329}}].

\bibitem{Graham:2015ouw}
P.~W. Graham, I.~G. Irastorza, S.~K. Lamoreaux, A.~Lindner, and K.~A. van
  Bibber, {\it {Experimental Searches for the Axion and Axion-Like Particles}},
   {\em Ann. Rev. Nucl. Part. Sci.} {\bf 65} (2015) 485--514,
  [\href{http://arxiv.org/abs/1602.00039}{{\tt arXiv:1602.00039}}].

\bibitem{Marsh:2015xka}
D.~J.~E. Marsh, {\it {Axion Cosmology}},  {\em Phys. Rept.} {\bf 643} (2016)
  1--79, [\href{http://arxiv.org/abs/1510.07633}{{\tt arXiv:1510.07633}}].

\bibitem{Irastorza:2018dyq}
I.~G. Irastorza and J.~Redondo, {\it {New experimental approaches in the search
  for axion-like particles}},  {\em Prog. Part. Nucl. Phys.} {\bf 102} (2018)
  89--159, [\href{http://arxiv.org/abs/1801.08127}{{\tt arXiv:1801.08127}}].

\bibitem{Raffelt:1990yz}
G.~G. Raffelt, {\it {Astrophysical methods to constrain axions and other novel
  particle phenomena}},  {\em Phys. Rept.} {\bf 198} (1990) 1--113.

\bibitem{Giannotti:2017hny}
M.~Giannotti, I.~G. Irastorza, J.~Redondo, A.~Ringwald, and K.~Saikawa, {\it
  {Stellar Recipes for Axion Hunters}},  {\em JCAP} {\bf 10} (2017) 010,
  [\href{http://arxiv.org/abs/1708.02111}{{\tt arXiv:1708.02111}}].

\bibitem{Dessert:2020lil}
C.~Dessert, J.~W. Foster, and B.~R. Safdi, {\it {X-ray Searches for Axions from
  Super Star Clusters}},  {\em Phys. Rev. Lett.} {\bf 125} (2020), no.~26
  261102, [\href{http://arxiv.org/abs/2008.03305}{{\tt arXiv:2008.03305}}].

\bibitem{Xiao:2020pra}
M.~Xiao, K.~M. Perez, M.~Giannotti, O.~Straniero, A.~Mirizzi, B.~W.
  Grefenstette, B.~M. Roach, and M.~Nynka, {\it {Constraints on Axionlike
  Particles from a Hard X-Ray Observation of Betelgeuse}},  {\em Phys. Rev.
  Lett.} {\bf 126} (2021), no.~3 031101,
  [\href{http://arxiv.org/abs/2009.09059}{{\tt arXiv:2009.09059}}].

\bibitem{Payez:2014xsa}
A.~Payez, C.~Evoli, T.~Fischer, M.~Giannotti, A.~Mirizzi, and A.~Ringwald, {\it
  {Revisiting the SN1987A gamma-ray limit on ultralight axion-like particles}},
   {\em JCAP} {\bf 02} (2015) 006, [\href{http://arxiv.org/abs/1410.3747}{{\tt
  arXiv:1410.3747}}].

\bibitem{Mirizzi:2007hr}
A.~Mirizzi, G.~G. Raffelt, and P.~D. Serpico, {\it {Signatures of Axion-Like
  Particles in the Spectra of TeV Gamma-Ray Sources}},  {\em Phys. Rev. D} {\bf
  76} (2007) 023001, [\href{http://arxiv.org/abs/0704.3044}{{\tt
  arXiv:0704.3044}}].

\bibitem{Hooper:2007bq}
D.~Hooper and P.~D. Serpico, {\it {Detecting Axion-Like Particles With Gamma
  Ray Telescopes}},  {\em Phys. Rev. Lett.} {\bf 99} (2007) 231102,
  [\href{http://arxiv.org/abs/0706.3203}{{\tt arXiv:0706.3203}}].

\bibitem{Hochmuth:2007hk}
K.~A. Hochmuth and G.~Sigl, {\it {Effects of Axion-Photon Mixing on Gamma-Ray
  Spectra from Magnetized Astrophysical Sources}},  {\em Phys. Rev. D} {\bf 76}
  (2007) 123011, [\href{http://arxiv.org/abs/0708.1144}{{\tt
  arXiv:0708.1144}}].

\bibitem{DeAngelis:2007wiw}
A.~De~Angelis, O.~Mansutti, and M.~Roncadelli, {\it {Axion-Like Particles,
  Cosmic Magnetic Fields and Gamma-Ray Astrophysics}},  {\em Phys. Lett. B}
  {\bf 659} (2008) 847--855, [\href{http://arxiv.org/abs/0707.2695}{{\tt
  arXiv:0707.2695}}].

\bibitem{DeAngelis:2007dqd}
A.~De~Angelis, M.~Roncadelli, and O.~Mansutti, {\it {Evidence for a new light
  spin-zero boson from cosmological gamma-ray propagation?}},  {\em Phys. Rev.
  D} {\bf 76} (2007) 121301, [\href{http://arxiv.org/abs/0707.4312}{{\tt
  arXiv:0707.4312}}].

\bibitem{Horns:2012kw}
D.~Horns, L.~Maccione, M.~Meyer, A.~Mirizzi, D.~Montanino, and M.~Roncadelli,
  {\it {Hardening of TeV gamma spectrum of AGNs in galaxy clusters by
  conversions of photons into axion-like particles}},  {\em Phys. Rev. D} {\bf
  86} (2012) 075024, [\href{http://arxiv.org/abs/1207.0776}{{\tt
  arXiv:1207.0776}}].

\bibitem{Berg:2016ese}
M.~Berg, J.~P. Conlon, F.~Day, N.~Jennings, S.~Krippendorf, A.~J. Powell, and
  M.~Rummel, {\it {Constraints on Axion-Like Particles from X-ray Observations
  of NGC1275}},  {\em Astrophys. J.} {\bf 847} (2017), no.~2 101,
  [\href{http://arxiv.org/abs/1605.01043}{{\tt arXiv:1605.01043}}].

\bibitem{Reynolds:2019uqt}
C.~S. Reynolds, M.~C.~D. Marsh, H.~R. Russell, A.~C. Fabian, R.~Smith,
  F.~Tombesi, and S.~Veilleux, {\it {Astrophysical limits on very light
  axion-like particles from Chandra grating spectroscopy of NGC 1275}},  {\em
  ApJ} {\bf 890} (7, 2019) [\href{http://arxiv.org/abs/1907.05475}{{\tt
  arXiv:1907.05475}}].

\bibitem{Marsh:2017yvc}
M.~C.~D. Marsh, H.~R. Russell, A.~C. Fabian, B.~P. McNamara, P.~Nulsen, and
  C.~S. Reynolds, {\it {A New Bound on Axion-Like Particles}},  {\em JCAP} {\bf
  12} (2017) 036, [\href{http://arxiv.org/abs/1703.07354}{{\tt
  arXiv:1703.07354}}].

\bibitem{Hook:2018iia}
A.~Hook, Y.~Kahn, B.~R. Safdi, and Z.~Sun, {\it {Radio Signals from Axion Dark
  Matter Conversion in Neutron Star Magnetospheres}},  {\em Phys. Rev. Lett.}
  {\bf 121} (2018), no.~24 241102, [\href{http://arxiv.org/abs/1804.03145}{{\tt
  arXiv:1804.03145}}].

\bibitem{Pshirkov:2007st}
M.~S. Pshirkov and S.~B. Popov, {\it {Conversion of Dark matter axions to
  photons in magnetospheres of neutron stars}},  {\em J. Exp. Theor. Phys.}
  {\bf 108} (2009) 384--388, [\href{http://arxiv.org/abs/0711.1264}{{\tt
  arXiv:0711.1264}}].

\bibitem{Huang:2018lxq}
F.~P. Huang, K.~Kadota, T.~Sekiguchi, and H.~Tashiro, {\it {Radio telescope
  search for the resonant conversion of cold dark matter axions from the
  magnetized astrophysical sources}},  {\em Phys. Rev. D} {\bf 97} (2018),
  no.~12 123001, [\href{http://arxiv.org/abs/1803.08230}{{\tt
  arXiv:1803.08230}}].

\bibitem{Safdi:2018oeu}
B.~R. Safdi, Z.~Sun, and A.~Y. Chen, {\it {Detecting Axion Dark Matter with
  Radio Lines from Neutron Star Populations}},  {\em Phys. Rev. D} {\bf 99}
  (2019), no.~12 123021, [\href{http://arxiv.org/abs/1811.01020}{{\tt
  arXiv:1811.01020}}].

\bibitem{Foster:2020pgt}
J.~W. Foster, Y.~Kahn, O.~Macias, Z.~Sun, R.~P. Eatough, V.~I. Kondratiev,
  W.~M. Peters, C.~Weniger, and B.~R. Safdi, {\it {Green Bank and Effelsberg
  Radio Telescope Searches for Axion Dark Matter Conversion in Neutron Star
  Magnetospheres}},  {\em Phys. Rev. Lett.} {\bf 125} (2020), no.~17 171301,
  [\href{http://arxiv.org/abs/2004.00011}{{\tt arXiv:2004.00011}}].

\bibitem{Darling:2020plz}
J.~Darling, {\it {Search for Axionic Dark Matter Using the Magnetar PSR
  J1745-2900}},  {\em Phys. Rev. Lett.} {\bf 125} (2020), no.~12 121103,
  [\href{http://arxiv.org/abs/2008.01877}{{\tt arXiv:2008.01877}}].

\bibitem{Battye:2021xvt}
R.~A. Battye, B.~Garbrecht, J.~I. McDonald, and S.~Srinivasan, {\it {Radio line
  properties of axion dark matter conversion in neutron stars}},  {\em JHEP}
  {\bf 09} (2021) 105, [\href{http://arxiv.org/abs/2104.08290}{{\tt
  arXiv:2104.08290}}].

\bibitem{Witte:2021arp}
S.~J. Witte, D.~Noordhuis, T.~D.~P. Edwards, and C.~Weniger, {\it {Axion-photon
  conversion in neutron star magnetospheres: The role of the plasma in the
  Goldreich-Julian model}},  {\em Phys. Rev. D} {\bf 104} (2021), no.~10
  103030, [\href{http://arxiv.org/abs/2104.07670}{{\tt arXiv:2104.07670}}].

\bibitem{Battye:2021yue}
R.~A. Battye, J.~Darling, J.~McDonald, and S.~Srinivasan, {\it {Towards Robust
  Constraints on Axion Dark Matter using PSR J1745-2900}},
  [\href{http://arxiv.org/abs/2107.01225}{{\tt arXiv:2107.01225}}].

\bibitem{Millar:2021gzs}
A.~J. Millar, S.~Baum, M.~Lawson, and M.~C.~D. Marsh, {\it {Axion-photon
  conversion in strongly magnetised plasmas}},  {\em JCAP} {\bf 11} (2021) 013,
  [\href{http://arxiv.org/abs/2107.07399}{{\tt arXiv:2107.07399}}].

\bibitem{Raffelt:1987im}
G.~Raffelt and L.~Stodolsky, {\it {Mixing of the Photon with Low Mass
  Particles}},  {\em Phys. Rev. D} {\bf 37} (1988) 1237.

\bibitem{Braaten:1993jw}
E.~Braaten and D.~Segel, {\it {Neutrino energy loss from the plasma process at
  all temperatures and densities}},  {\em Phys. Rev. D} {\bf 48} (1993)
  1478--1491, [\href{http://arxiv.org/abs/hep-ph/9302213}{{\tt
  hep-ph/9302213}}].

\bibitem{Dobrynina:2014qba}
A.~Dobrynina, A.~Kartavtsev, and G.~Raffelt, {\it {Photon-photon dispersion of
  TeV gamma rays and its role for photon-ALP conversion}},  {\em Phys. Rev. D}
  {\bf 91} (2015) 083003, [\href{http://arxiv.org/abs/1412.4777}{{\tt
  arXiv:1412.4777}}]. [Erratum: Phys.Rev.D 95, 109905 (2017)].

\bibitem{companion}
K.~Bondarenko, A.~Boyarsky, J.~Pradler, and A.~Sokolenko, {\it {}},
  [\href{http://arxiv.org/abs/in preparation}{{\tt in preparation}}].

\bibitem{Sarazin:1986zz}
C.~L. Sarazin, {\it {X-ray emission from clusters of galaxies}},  {\em Rev.
  Mod. Phys.} {\bf 58} (1986) 1--115.

\bibitem{Garcia:2020qrp}
A.~A. Garcia, K.~Bondarenko, S.~Ploeckinger, J.~Pradler, and A.~Sokolenko, {\it
  {Effective photon mass and (dark) photon conversion in the inhomogeneous
  Universe}},  {\em JCAP} {\bf 10} (2020) 011,
  [\href{http://arxiv.org/abs/2003.10465}{{\tt arXiv:2003.10465}}].

\bibitem{Sikivie:1983ip}
P.~Sikivie, {\it {Experimental Tests of the Invisible Axion}},  {\em Phys. Rev.
  Lett.} {\bf 51} (1983) 1415--1417. [Erratum: Phys.Rev.Lett. 52, 695 (1984)].

\bibitem{Sikivie:1985yu}
P.~Sikivie, {\it {Detection Rates for 'Invisible' Axion Searches}},  {\em Phys.
  Rev. D} {\bf 32} (1985) 2988. [Erratum: Phys.Rev.D 36, 974 (1987)].

\bibitem{Grossman:2002by}
Y.~Grossman, S.~Roy, and J.~Zupan, {\it {Effects of initial axion production
  and photon axion oscillation on type Ia supernova dimming}},  {\em Phys.
  Lett. B} {\bf 543} (2002) 23--28,
  [\href{http://arxiv.org/abs/hep-ph/0204216}{{\tt hep-ph/0204216}}].

\bibitem{Lai:2001di}
D.~Lai and W.~C.~G. Ho, {\it {Resonant conversion of photon modes due to vacuum
  polarization in a magnetized plasma: implications for x-ray emission from
  magnetars}},  {\em Astrophys. J.} {\bf 566} (2002) 373,
  [\href{http://arxiv.org/abs/astro-ph/0108127}{{\tt astro-ph/0108127}}].

\bibitem{Lai:2006af}
D.~Lai and J.~Heyl, {\it {Probing Axions with Radiation from Magnetic Stars}},
  {\em Phys. Rev. D} {\bf 74} (2006) 123003,
  [\href{http://arxiv.org/abs/astro-ph/0609775}{{\tt astro-ph/0609775}}].

\bibitem{Goldreich1969}
P.~{Goldreich} and W.~H. {Julian}, {\it {Pulsar Electrodynamics}},  {\em \apj}
  {\bf 157} (Aug., 1969) 869.

\bibitem{Sobyanin:2016acr}
D.~N. Sob'yanin, {\it {Breakdown of the Goldreich-Julian Relation in a Neutron
  Star}},  {\em Astron. Lett.} {\bf 42} (2016) 745,
  [\href{http://arxiv.org/abs/1612.09139}{{\tt arXiv:1612.09139}}].

\bibitem{Lyutikov:2007fn}
M.~Lyutikov, {\it {Neutron star magnetospheres: The binary pulsar, Crab and
  magnetars}},  {\em AIP Conf. Proc.} {\bf 968} (2008), no.~1 77--84,
  [\href{http://arxiv.org/abs/0708.1024}{{\tt arXiv:0708.1024}}].

\bibitem{Timokhin:2015dua}
A.~N. Timokhin and A.~K. Harding, {\it {On the polar cap cascade pair
  multiplicity of young pulsars}},  {\em Astrophys. J.} {\bf 810} (2015), no.~2
  144, [\href{http://arxiv.org/abs/1504.02194}{{\tt arXiv:1504.02194}}].

\bibitem{Cruz:2020vfm}
F.~Cruz, T.~Grismayer, and L.~O. Silva, {\it {Kinetic model of large-amplitude
  oscillations in neutron star pair cascades}},  {\em Astrophys. J.} {\bf 908}
  (2021), no.~2 149, [\href{http://arxiv.org/abs/2012.05587}{{\tt
  arXiv:2012.05587}}].

\bibitem{1998PhRvE..57.3399G}
M.~{Gedalin}, D.~B. {Melrose}, and E.~{Gruman}, {\it {Long waves in a
  relativistic pair plasma in a strong magnetic field}},  {\em \pre} {\bf 57}
  (Mar., 1998) 3399--3410.

\bibitem{Leroy:2019ghm}
M.~Leroy, M.~Chianese, T.~D.~P. Edwards, and C.~Weniger, {\it {Radio Signal of
  Axion-Photon Conversion in Neutron Stars: A Ray Tracing Analysis}},  {\em
  Phys. Rev. D} {\bf 101} (2020), no.~12 123003,
  [\href{http://arxiv.org/abs/1912.08815}{{\tt arXiv:1912.08815}}].

\bibitem{Kaspi:2017fwg}
V.~M. Kaspi and A.~Beloborodov, {\it {Magnetars}},  {\em Ann. Rev. Astron.
  Astrophys.} {\bf 55} (2017) 261--301,
  [\href{http://arxiv.org/abs/1703.00068}{{\tt arXiv:1703.00068}}].

\bibitem{Tsai:1974fa}
W.-y. Tsai and T.~Erber, {\it {Photon Pair Creation in Intense Magnetic
  Fields}},  {\em Phys. Rev. D} {\bf 10} (1974) 492.

\bibitem{Tsai:1975iz}
W.-y. Tsai and T.~Erber, {\it {The Propagation of Photons in Homogeneous
  Magnetic Fields: Index of Refraction}},  {\em Phys. Rev. D} {\bf 12} (1975)
  1132.

\bibitem{Philippov:2014mqa}
A.~A. Philippov, A.~Spitkovsky, and B.~Cerutti, {\it {Ab-initio pulsar
  magnetosphere: three-dimensional particle-in-cell simulations of oblique
  pulsars}},  {\em Astrophys. J. Lett.} {\bf 801} (2015), no.~1 L19,
  [\href{http://arxiv.org/abs/1412.0673}{{\tt arXiv:1412.0673}}].

\bibitem{Turolla:2015mwa}
R.~Turolla, S.~Zane, and A.~Watts, {\it {Magnetars: the physics behind
  observations. A review}},  {\em Rept. Prog. Phys.} {\bf 78} (2015), no.~11
  116901, [\href{http://arxiv.org/abs/1507.02924}{{\tt arXiv:1507.02924}}].

\bibitem{Carenza:2023nck}
P.~Carenza and M.~C.~D. Marsh, {\it {On the applicability of the Landau-Zener
  formula to axion-photon conversion}},  {\em JCAP} {\bf 04} (2023) 021,
  [\href{http://arxiv.org/abs/2302.02700}{{\tt arXiv:2302.02700}}].

\bibitem{Torne:2015rha}
P.~Torne et~al., {\it {Simultaneous multifrequency radio observations of the
  Galactic Centre magnetar SGR J1745\ensuremath{-}2900}},  {\em Mon. Not. Roy.
  Astron. Soc.} {\bf 451} (2015), no.~1 L50--L54,
  [\href{http://arxiv.org/abs/1504.07241}{{\tt arXiv:1504.07241}}].

\bibitem{Torne:2016zid}
P.~Torne, G.~Desvignes, R.~P. Eatough, R.~Karuppusamy, G.~Paubert, M.~Kramer,
  I.~Cognard, D.~J. Champion, and L.~G. Spitler, {\it {Detection of the
  magnetar SGR J1745\ensuremath{-}2900 up to 291 GHz with evidence of polarized
  millimetre emission}},  {\em Mon. Not. Roy. Astron. Soc.} {\bf 465} (2017),
  no.~1 242--247, [\href{http://arxiv.org/abs/1610.07616}{{\tt
  arXiv:1610.07616}}].

\bibitem{Kennea:2013dfa}
J.~A. Kennea et~al., {\it {Swift Discovery of a New Soft Gamma Repeater, SGR
  J1745-29, near Sagittarius A*}},  {\em Astrophys. J. Lett.} {\bf 770} (2013)
  L24, [\href{http://arxiv.org/abs/1305.2128}{{\tt arXiv:1305.2128}}].

\bibitem{Mori:2013yda}
K.~Mori et~al., {\it {NuSTAR discovery of a 3.76-second transient magnetar near
  Sagittarius A*}},  {\em Astrophys. J. Lett.} {\bf 770} (2013) L23,
  [\href{http://arxiv.org/abs/1305.1945}{{\tt arXiv:1305.1945}}].

\bibitem{Beloborodov:2008df}
A.~M. Beloborodov, {\it {Untwisting magnetospheres of neutron stars}},  {\em
  Astrophys. J.} {\bf 703} (2009) 1044--1060,
  [\href{http://arxiv.org/abs/0812.4873}{{\tt arXiv:0812.4873}}].

\bibitem{githublimits}
C.~O'Hare, ``Axionlimits.'' \url{https://github.com/cajohare/AxionLimits},
  2021.

\bibitem{Dessert:2022yqq}
C.~Dessert, D.~Dunsky, and B.~R. Safdi, {\it {Upper limit on the axion-photon
  coupling from magnetic white dwarf polarization}},  {\em Phys. Rev. D} {\bf
  105} (2022), no.~10 103034, [\href{http://arxiv.org/abs/2203.04319}{{\tt
  arXiv:2203.04319}}].

\bibitem{Noordhuis:2022ljw}
D.~Noordhuis, A.~Prabhu, S.~J. Witte, A.~Y. Chen, F.~Cruz, and C.~Weniger, {\it
  {Novel Constraints on Axions Produced in Pulsar Polar-Cap Cascades}},
  [\href{http://arxiv.org/abs/2209.09917}{{\tt arXiv:2209.09917}}].

\bibitem{CAST:2017uph}
{\bf CAST} Collaboration, V.~Anastassopoulos et~al., {\it {New CAST Limit on
  the Axion-Photon Interaction}},  {\em Nature Phys.} {\bf 13} (2017) 584--590,
  [\href{http://arxiv.org/abs/1705.02290}{{\tt arXiv:1705.02290}}].

\bibitem{Gramolin:2020ict}
A.~V. Gramolin, D.~Aybas, D.~Johnson, J.~Adam, and A.~O. Sushkov, {\it {Search
  for axion-like dark matter with ferromagnets}},  {\em Nature Phys.} {\bf 17}
  (2021), no.~1 79--84, [\href{http://arxiv.org/abs/2003.03348}{{\tt
  arXiv:2003.03348}}].

\bibitem{Salemi:2021gck}
C.~P. Salemi et~al., {\it {The search for low-mass axion dark matter with
  ABRACADABRA-10cm}},  [\href{http://arxiv.org/abs/2102.06722}{{\tt
  arXiv:2102.06722}}].

\bibitem{Ortiz:2020tgs}
M.~D. Ortiz et~al., {\it {Design of the ALPS II optical system}},  {\em Phys.
  Dark Univ.} {\bf 35} (2022) 100968,
  [\href{http://arxiv.org/abs/2009.14294}{{\tt arXiv:2009.14294}}].

\bibitem{2013ITAS...23T0604S}
I.~{Shilon}, A.~{Dudarev}, H.~{Silva}, and H.~H.~J. {ten Kate}, {\it
  {Conceptual Design of a New Large Superconducting Toroid for IAXO, the New
  International AXion Observatory}},  {\em IEEE Transactions on Applied
  Superconductivity} {\bf 23} (June, 2013) 4500604--4500604,
  [\href{http://arxiv.org/abs/1212.4633}{{\tt arXiv:1212.4633}}].

\bibitem{Michimura:2019qxr}
Y.~Michimura, Y.~Oshima, T.~Watanabe, T.~Kawasaki, H.~Takeda, M.~Ando,
  K.~Nagano, I.~Obata, and T.~Fujita, {\it {DANCE: Dark matter Axion search
  with riNg Cavity Experiment}},  {\em J. Phys. Conf. Ser.} {\bf 1468} (2020),
  no.~1 012032, [\href{http://arxiv.org/abs/1911.05196}{{\tt
  arXiv:1911.05196}}].

\bibitem{Liu:2018icu}
H.~Liu, B.~D. Elwood, M.~Evans, and J.~Thaler, {\it {Searching for Axion Dark
  Matter with Birefringent Cavities}},  {\em Phys. Rev. D} {\bf 100} (2019),
  no.~2 023548, [\href{http://arxiv.org/abs/1809.01656}{{\tt
  arXiv:1809.01656}}].

\bibitem{PhysRevLett.59.839}
S.~DePanfilis, A.~C. Melissinos, B.~E. Moskowitz, J.~T. Rogers, Y.~K.
  Semertzidis, W.~U. Wuensch, H.~J. Halama, A.~G. Prodell, W.~B. Fowler, and
  F.~A. Nezrick, {\it Limits on the abundance and coupling of cosmic axions at
  $4.5<{m}_{a}<5.0$ \ensuremath{\mu}ev},  {\em Phys. Rev. Lett.} {\bf 59} (Aug,
  1987) 839--842.

\bibitem{HAYSTAC:2018rwy}
{\bf HAYSTAC} Collaboration, L.~Zhong et~al., {\it {Results from phase 1 of the
  HAYSTAC microwave cavity axion experiment}},  {\em Phys. Rev. D} {\bf 97}
  (2018), no.~9 092001, [\href{http://arxiv.org/abs/1803.03690}{{\tt
  arXiv:1803.03690}}].

\bibitem{ADMX:2019uok}
{\bf ADMX} Collaboration, T.~Braine et~al., {\it {Extended Search for the
  Invisible Axion with the Axion Dark Matter Experiment}},  {\em Phys. Rev.
  Lett.} {\bf 124} (2020), no.~10 101303,
  [\href{http://arxiv.org/abs/1910.08638}{{\tt arXiv:1910.08638}}].

\bibitem{HAYSTAC:2020kwv}
{\bf HAYSTAC} Collaboration, K.~M. Backes et~al., {\it {A quantum-enhanced
  search for dark matter axions}},  {\em Nature} {\bf 590} (2021), no.~7845
  238--242, [\href{http://arxiv.org/abs/2008.01853}{{\tt arXiv:2008.01853}}].

\bibitem{Jeong:2020cwz}
J.~Jeong, S.~Youn, S.~Bae, J.~Kim, T.~Seong, J.~E. Kim, and Y.~K. Semertzidis,
  {\it {Search for Invisible Axion Dark Matter with a Multiple-Cell
  Haloscope}},  {\em Phys. Rev. Lett.} {\bf 125} (2020), no.~22 221302,
  [\href{http://arxiv.org/abs/2008.10141}{{\tt arXiv:2008.10141}}].

\bibitem{Lee:2020cfj}
S.~Lee, S.~Ahn, J.~Choi, B.~R. Ko, and Y.~K. Semertzidis, {\it {Axion Dark
  Matter Search around 6.7 $\mu$eV}},  {\em Phys. Rev. Lett.} {\bf 124} (2020),
  no.~10 101802, [\href{http://arxiv.org/abs/2001.05102}{{\tt
  arXiv:2001.05102}}].

\bibitem{ADMX:2021nhd}
{\bf ADMX} Collaboration, C.~Bartram et~al., {\it {Search for Invisible Axion
  Dark Matter in the 3.3\textendash{}4.2\,\,\ensuremath{\mu}eV Mass Range}},
  {\em Phys. Rev. Lett.} {\bf 127} (2021), no.~26 261803,
  [\href{http://arxiv.org/abs/2110.06096}{{\tt arXiv:2110.06096}}].

\bibitem{Escudero:2023vgv}
M.~Escudero, C.~K. Pooni, M.~Fairbairn, D.~Blas, X.~Du, and D.~J.~E. Marsh,
  {\it {Axion Star Explosions: A New Source for Axion Indirect Detection}},
  [\href{http://arxiv.org/abs/2302.10206}{{\tt arXiv:2302.10206}}].

\bibitem{Kramer:2006ha}
M.~Kramer, A.~G. Lyne, J.~T. O'Brien, C.~A. Jordan, and D.~R. Lorimer, {\it {A
  periodically active pulsar giving insight into magnetospheric physics}},
  {\em Science} {\bf 312} (2006) 549--551,
  [\href{http://arxiv.org/abs/astro-ph/0604605}{{\tt astro-ph/0604605}}].

\bibitem{Watts:2014trg}
A.~Watts et~al., {\it {Understanding the Neutron Star Population with the
  SKA}},  {\em PoS} {\bf AASKA14} (2015) 039,
  [\href{http://arxiv.org/abs/1501.00005}{{\tt arXiv:1501.00005}}].

\bibitem{Keane:2014vja}
E.~F. Keane et~al., {\it {A Cosmic Census of Radio Pulsars with the SKA}},
  {\em PoS} {\bf AASKA14} (2015) 040,
  [\href{http://arxiv.org/abs/1501.00056}{{\tt arXiv:1501.00056}}].

\bibitem{Karastergiou:2014cka}
A.~Karastergiou et~al., {\it {Understanding pulsar magnetospheres with the
  SKA}},  {\em PoS} {\bf AASKA14} (2015) 038,
  [\href{http://arxiv.org/abs/1501.00126}{{\tt arXiv:1501.00126}}].

\bibitem{Antoniadis:2015nga}
J.~Antoniadis, L.~Guillemot, A.~Possenti, S.~Bogdanov, J.~Gelfand, M.~Kramer,
  R.~Mignani, B.~Stappers, and P.~Torne, {\it {Multi-wavelength,
  Multi-Messenger Pulsar Science in the SKA Era}},  {\em PoS} {\bf AASKA14}
  (2015) 157, [\href{http://arxiv.org/abs/1501.05591}{{\tt arXiv:1501.05591}}].

\bibitem{Chu:2021ktf}
C.-Y. Chu, C.~Y. Ng, A.~K.~H. Kong, and H.-K. Chang, {\it {High-frequency radio
  observations of two magnetars, PSR J1622 \ensuremath{-} 4950 and 1E 1547.0
  \ensuremath{-} 5408}},  {\em Mon. Not. Roy. Astron. Soc.} {\bf 503} (2021),
  no.~1 1214--1220, [\href{http://arxiv.org/abs/2102.02466}{{\tt
  arXiv:2102.02466}}].

\bibitem{Torne:2021yad}
P.~Torne et~al., {\it {Searching for pulsars in the Galactic centre at 3 and 2
  mm}},  {\em Astron. Astrophys.} {\bf 650} (2021) A95,
  [\href{http://arxiv.org/abs/2103.16581}{{\tt arXiv:2103.16581}}].

\bibitem{Torne:2022xfo}
P.~Torne et~al., {\it {Submillimeter Pulsations from the Magnetar XTE
  J1810-197}},  {\em Astrophys. J. Lett.} {\bf 925} (2022), no.~2 L17,
  [\href{http://arxiv.org/abs/2201.07820}{{\tt arXiv:2201.07820}}].

\bibitem{Petri:2016tqe}
J.~P\'etri, {\it {Theory of pulsar magnetosphere and wind}},  {\em J. Plasma
  Phys.} {\bf 82} (2016), no.~5 635820502,
  [\href{http://arxiv.org/abs/1608.04895}{{\tt arXiv:1608.04895}}].

\bibitem{Chen:2016qkk}
A.~Y. Chen and A.~M. Beloborodov, {\it {Particle-in-cell simulations of the
  twisted magnetospheres of magnetars. I}},  {\em Astrophys. J.} {\bf 844}
  (2017), no.~2 133, [\href{http://arxiv.org/abs/1610.10036}{{\tt
  arXiv:1610.10036}}].

\bibitem{Cerutti:2016ttn}
B.~Cerutti and A.~Beloborodov, {\it {Electrodynamics of pulsar
  magnetospheres}},  {\em Space Sci. Rev.} {\bf 207} (2017), no.~1-4 111--136,
  [\href{http://arxiv.org/abs/1611.04331}{{\tt arXiv:1611.04331}}].

\bibitem{Brambilla:2017puc}
G.~Brambilla, C.~Kalapotharakos, A.~Timokhin, A.~Harding, and D.~Kazanas, {\it
  {Electron\textendash{}Positron Pair Flow and Current Composition in the
  Pulsar Magnetosphere}},  {\em Astrophys. J.} {\bf 858} (2018), no.~2 81,
  [\href{http://arxiv.org/abs/1710.03536}{{\tt arXiv:1710.03536}}].

\bibitem{Carrasco:2020sxg}
F.~Carrasco and M.~Shibata, {\it {Magnetosphere of an orbiting neutron star}},
  {\em Phys. Rev. D} {\bf 101} (2020), no.~6 063017,
  [\href{http://arxiv.org/abs/2001.04210}{{\tt arXiv:2001.04210}}].

\bibitem{2020ApJ...889...69C}
A.~Y. {Chen}, F.~{Cruz}, and A.~{Spitkovsky}, {\it {Filling the Magnetospheres
  of Weak Pulsars}},  {\em \apj} {\bf 889} (Jan., 2020) 69,
  [\href{http://arxiv.org/abs/1911.00059}{{\tt arXiv:1911.00059}}].

\end{thebibliography}\endgroup

\newpage

\appendix

\onecolumn

\begin{center}
    \textbf{Supplemental Material}
\end{center}

\section{Resonance condition in strong magnetic field}

We start from Maxwell's equations supplied with the external current  $\vec J = g_{a\gamma}  \vec B_0 \partial_t a$ where $\vec B_0$ is the external magnetic field,
\begin{align}
    \nabla \times \vec E + \partial_t \vec B &= 0, \quad \vec D = \bm\epsilon\cdot \vec E, \\
    \nabla\times \vec H - \partial_t \vec D & =  g_{a\gamma}  \vec B_0 \partial_t a, \quad  \vec B  = \bm \mu \vec H.
\end{align}
Here, $\bm \epsilon$ and $\bm \mu$ are the dielectric and magnetic permeability tensors. 
In a coordinate system in which 
the external $\vec B_0$ field is in $z$-direction, the dielectric and magnetic permeability tensors take the following form, 
\begin{align}
\bm\epsilon_{\vec B_0\parallel \hat z} =
\begin{pmatrix}
1 & 0 & 0 \\
0 & 1 & 0 \\
0 & 0 & 1-\omega_p^2/\omega^2
\end{pmatrix} - 
A
\begin{pmatrix}
2   \vec B_0^2 & 0 & 0 \\
0 & 2  \vec B_0^2 & 0 \\
0 & 0 &    -5 \vec B_0^2
\end{pmatrix}. 
\end{align}
The first contribution is the plasma effect, where we neglect relativistic corrections. As we shall see below, our results are robust to restricting ourselves to spatial regions that are not in the immediate vicinity of the neutron star surface where relativistic electron populations occur. The second contribution is  the vacuum polarization contribution to the dielectric tensor with  $A = 4\alpha^2/45 m_e^4$.

 In addition, the magnetic permeability tensor reads  
$
\bm\mu=
(1+2 A B_0^2) \bm 1_{3\times 3} .
$

Taking the time dependence of the axion and electric field as $e^{-i\omega t}$, and neglecting any time derivatives of $\vec B_0$, $\bm \epsilon$ and $\mu$ the Maxwell equations can be combined. Supplied with the Klein-Gordan equation for the axion, the equations to solve are,
\begin{align}
-\omega^2(\boldsymbol{\epsilon} \cdot \vec{E}) 
+ \nabla \times [\bm \mu^{-1} (\nabla \times \vec E) ]
& =g_{a \gamma }\omega^2 \vec{B}_{0}  a , \\
\left(-\omega^2-\nabla^{2}+m_{a}^{2}\right) a & =g_{a \gamma } \vec{E} \cdot \vec{B}_{0} ,
\end{align}

If we now switch to a coordinate system where the photon propagates in the $z$-direction, magnetic field has angle $\theta$ to the photon direction and is in $xz$-plane, that requires a rotation of the magnetic field and the dielectric tensor as
\begin{align*}
\vec B_0 = B_0 R(\theta) \hat z = B_0 \begin{pmatrix}
    \sin\theta \\ 0 \\ \cos\theta
\end{pmatrix}, \quad
\bm\epsilon=R(\theta) \cdot \bm\epsilon_{\vec B_0\parallel \hat z} \cdot R(-\theta) ,
\end{align*}
with the rotation matrix given by
\begin{align*}
R(\theta)=\left(\begin{array}{ccc}
\cos \theta & 0 & \sin \theta \\
0 & 1 & 0 \\
-\sin \theta & 0 & \cos \theta
\end{array}\right) .
\end{align*}
Since $\bm \mu$ is diagonal, it remains invariant under the rotation and its inverse is simply given by $\bm \mu ^{-1} =(1+2 A B_0^2)^{-1} \bm 1_{3\times 3} $. 
The expression for the dielectric tensor becomes,
\begin{align*}
\bm\epsilon=\begin{pmatrix}
1 - \sin^2\theta (\omega_p^2/\omega^2) 
& 0 &  - \sin\theta\cos\theta (\omega_p^2/\omega^2) \\
0 & 1 & 0 \\
- \sin\theta\cos\theta (\omega_p^2/\omega^2) & 0 & 1 - \cos^2\theta (\omega_p^2/\omega^2)
\end{pmatrix} 
+
\frac{1}{2}A B_0^2
\begin{pmatrix}
3-7 \cos 2\theta 
& 0 &   14 \sin\theta\cos\theta  \\
0 & -4 & 0 \\
14 \sin\theta\cos\theta  & 0 & 3+7 \cos 2\theta 
\end{pmatrix} 
.
\end{align*}

We see that $E_y$ is not directly coupled to the axion since $ B_{0y}=0$ and it is only generated through further spatial derivatives, which makes it  suppressed. In the following we  hence treat the reduced problem with the $y$-components neglected.
If we are in addition to neglecting $E_y$ also neglect any second order derivatives or products of first order derivatives that do not include $\partial_z$, we arrive at the following equations
\begin{align}
  -\omega^2 (\epsilon_{xx}E_x + \epsilon_{xz}E_z) - \mu_{yy}^{-1} ( \partial_z^2 E_x - \partial_x\partial_z E_z ) - (\partial_z \mu_{yy}^{-1}) (\partial_z E_x - \partial_x E_z )
  & = g_{a\gamma} \omega^2 B_{0x} a , \\
  -\omega^2 (\epsilon_{zx}E_x + \epsilon_{zz}E_z) 
  + \mu_{yy}^{-1} \partial_x\partial_z E_x 
+ \partial_x \mu_{yy}^{-1} \partial_z E_x  
  & = g_{a\gamma} \omega^2 B_{0z} a , 
\end{align} 
where $\mu_{xx} = (\bm \mu^{-1})_{xx}$ and so forth.
We may now use the second equation to express $E_z$
\begin{align}
    E_z = - \frac{\epsilon_{zx}}{\epsilon_{zz}}E_x  + \frac{1}{\epsilon_{zz}\omega^2 } \left[  - g_{a\gamma} \omega^2 B_{0z} a  + \mu_{yy}^{-1}\partial_x\partial_z E_x + (\partial_x \mu_{yy}^{-1}) \partial_z E_x \right] .
\end{align}

The resulting expression is rather cumbersome. Following~\cite{Millar:2021gzs} in a number of approximations, neglecting any external electric fields as well as any derivatives of $B_0^2$ (which also amounts to neglecting derivatives of $\mu$) then we arrive at the following expression in a generalization of~\cite{Millar:2021gzs},
\begin{align}
 - \partial^2_z E_x + &
 \frac{(\omega_p^2 -7 A B_0^2 \omega ^2 ) \sin 2 \theta }{ \omega ^2 - \omega_p^2 \cos ^2\theta +\frac{1}{2} A B_0^2 \omega ^2 (7
   \cos 2 \theta +3)
   } \partial_x\partial_z E_x
   +\frac{\omega ^2 \omega_p (1 - 2 A
   B_0^2)  \sin 2 \theta  }{\left[\omega ^2-\omega_p^2 \cos ^2\theta + \frac{1}{2} A B_0^2 \omega ^2 (7 \cos 2
   \theta +3)
   \right]^2} \left( \partial_z E_x \partial_x\omega_p+\partial_xE_x \partial_z\omega_p\right)
    \\
  & =\frac{(1 - 4 A^2 B_0^4) \omega^2  \left(\omega ^2 - \omega_p^2 + 5
   A B_0^2 \omega ^2 \right)}{\omega^2 - \omega_p^2 \cos ^2\theta
   + \frac{1}{2} A B_0^2 \omega ^2 (7 \cos 2
   \theta +3) } E_x
   +B_0 g_{a\gamma} a  
  \frac{(1 - 4 A^2 B_0^4)  \omega ^4 \sin \theta }{\omega ^2-\omega_p^2 \cos
   ^2\theta + \frac{1}{2} A B_0^2 \omega ^2 (7 \cos 2
   \theta +3)} .
\end{align}
The factors $(1 - 4 A^2 B_0^4)$ are of higher order, originating from the the product of elements of $\bm \epsilon$ and $\bm \mu^{-1}$ and we will neglect them.
We now consider the propagation in the $z$-direction and an ansatz for the envelopes of the electric and axion fields (time dependence as  introduced above) 
\begin{align}
E_x \equiv \tilde{E}_x(x, z) e^{i(\omega t-k z)}, \quad a \equiv \tilde{a}(z) e^{i(\omega t-k z)}.
\end{align}
Following the WKB approximation detailed in~\cite{Millar:2021gzs} 
we arrive at
\begin{align}
\label{eom}
& 2 i k \partial_z \tilde{E}_x   - 2 i k \frac{(\omega_p^2 -7 A\omega^2 B_0^2 ) \xi}{\omega^2 \tan \theta} \partial_x \tilde{E}_x   \simeq 
\left[m_a^2 
- \omega^2 + \xi \csc^2\theta (\omega^2-\omega_p^2+5A\omega^2 B_0^2)
+ i k \mathcal{D}\right] \tilde{E}_x +  \frac{\omega^2 \xi}{\sin \theta} g_{a \gamma} \tilde{a} B_{\mathrm{NS}} 
\end{align}
Here we defined, 
\begin{align}
\xi \equiv \frac{\omega^2 \sin ^2 \theta}{\omega^2-\omega_p^2 \cos ^2 \theta + \frac{1}{2} A B_0^2 \omega ^2 (7 \cos 2
   \theta +3) }, \quad \mathcal{D} = \frac{2\omega_p\xi^2 (1-2AB_0^2)}{\omega^2 \tan\theta \sin^2\theta} \partial_x \omega_p .
\end{align}
Defining the differential operator 
\begin{align}
\partial_s =\partial_z -\frac{(\omega_p^2 -7 A\omega^2 B_0^2 )\xi }{\omega^2 \tan \theta}\partial_x ,
\end{align}
we obtain the Schroedinger type equation, 
\begin{align}
    i \partial_s \tilde E_x = 
\frac{1}{2 k} \left\{ \left[m_a^2 
- \omega^2 + \xi \csc^2\theta (\omega^2-\omega_p^2+5A\omega^2 B_0^2)
+ i k \mathcal{D}\right] \tilde{E}_x +  \frac{\omega^2 \xi}{\sin \theta} g_{a \gamma} \tilde{a} B_{\mathrm{NS}}  \right\}
\end{align}
This equation can be solved in the stationary phase approximation which yields the resonance condition.
Neglecting an overall phase, the solution reads, 
\begin{align}
    i \tilde E_x(s) = & \frac{1}{2k}\int_0^s ds'\  \frac{\omega^2 \xi}{\sin \theta} g_{a \gamma} \tilde{a} B_{\mathrm{NS}} \exp\left\{\frac{1}{2}\int_0^{s'} ds'' \mathcal{D}\right\} 
    \exp\left\{
    -i \int_0^{s'} ds''\ 
    \frac{1}{2k}\left[m_a^2
- \omega^2 + \xi \csc^2\theta (\omega^2-\omega_p^2+5A\omega^2 B_0^2)
\right] 
    \right\}
\end{align}
Let us first solve for $A=0$, i.e., without vacuum polarization. Then
\begin{align}
    f(s) = - \int_0^s \frac{1}{2k}\left[m_a^2-\xi \omega_p^2 \right] \quad \Rightarrow \quad  \partial_s f(s) = -  \frac{1}{2k}\left[m_a^2-\xi \omega_p^2 \right] \stackrel{s=s_0} = 0 \quad (A=0)
\end{align}
The latter relation gives the stationary point $s_0$ with the resonance condition, $m_a^2=\xi \omega_p^2$ reproducing the expression~\cite{Millar:2021gzs},
\begin{align}
{\omega}_p^2=\frac{m_a^2 \omega^2}{m_a^2 \cos ^2 \theta+\omega^2 \sin ^2 \theta} = 
\frac{2 m_a^2 \omega^2}{m_a^2  + \omega^2 + (m_a^2 -\omega^2) \cos 2\theta}\quad (A=0)
\end{align}
In the current context, this is being replaced by the following resonance condition,
\begin{align}
    m_a^2 \simeq \omega ^2 - 
   \left[\frac{\cos ^2\theta}{\omega^2 \left(1-2 A B_0^2\right)}+\frac{\sin
   ^2\theta}{ \omega^2(1+5 A B_0^2)-\omega_p^2}\right]^{-1}
   \label{eq:modified_res_condition}
\end{align}
For the special case of $\theta = \pi/2$ this becomes, 
\begin{align}
     m_a^2 \simeq \omega_p ^2 - \frac{4\alpha^2\omega^2 B_0^2}{9 m_e^4 } .
     \label{eq:res_cond_pi/2}
\end{align}
Comparing the negative contribution on the right-hand side with the known correction to the effective photon mass~\cite{Dobrynina:2014qba} $m_{\rm eff}^2|_{\rm vac.pol.} = - (44/135)  \alpha^2\omega^2 B_0^2/m_e^4 $ we see that the numerical factor is different by $35\%$. This difference is expected, because the negative contribution to the photon mass in~\cite{Dobrynina:2014qba} was calculated for the isotropic case, where electrons are free to move in all three spatial directions, while in the anisotropic case considered here, electrons  only move along magnetic field lines. Because of this, we retain an angular dependence in Eq.~\eqref{eq:modified_res_condition} for the resonant condition.

An important question in relation to our work is to what extend a detailed treatment such as the one above modifies the simple assumptions taking in the paper. For obtaining an answer, let us consider  $m_a \ll \omega_p$, such that resonant conversion is only possible if a cancellation between plasma frequency and the negative contribution from the magnetic field happens (termed ``double lens effect'' in the main paper). Neglecting the $m_a^2$ term in \eqref{eq:modified_res_condition}, one can rewrite the equation as
\begin{equation}
    \omega_p^2 \simeq (5 - 2 \cot^2 \theta) A \omega^2 B_0^2.
    \label{eq:res_cond_with_theta}
\end{equation}
We see that such cancellation is possible if $\cot^2 \theta < 5/2$ or
\begin{equation}
    \theta_{\text{cr}} < \theta < 180^{\circ} - \theta_{\text{cr}}, \qquad  \theta_{\text{cr}} = \arctan\left(\sqrt{2/5}\right) \approx 32.3^{\circ}.
\end{equation}
Hence, in a wide range of $\theta$ the resonance condition is approximately given by~\eqref{eq:res_cond_pi/2}. For example, the condition~\eqref{eq:res_cond_with_theta} coincides with~\eqref{eq:res_cond_pi/2} with a precision of $20\%$ if $55^{\circ}<\theta< 125^{\circ}$.


\section{Influence of the place of photon creation}

One of the important assumptions of our paper was the creation of the radio photon close to the surface of the magnetar. In this section, we analyze the influence of this assumption on the sensitivity region of the axion-photon conversion (Fig. 3 in the main text) and find that the dependence of the lower bound of this region on the radius of the radio signal creation is surprisingly weak (for $m_a^2 \ll \omega_p^2$).

In this section we will use the same assumption as in the main paper: the magnitude of the magnetic field decays as a power law,
\begin{equation}
    B(r) = B_0 \left( \frac{r_0}{r}\right)^3, 
\end{equation}
we will use average values for trigonometric function (assuming isotropic distribution), and electron number density is given by Goldreich-Julian model, $n_e (r) = \Omega B(r)/e$.

The experimental signature that we found from the double-lens effect is a kink in the spectrum of the radio waves. It appears when $B_+$ solution of the resonance condition is equal to the maximal value $B_{\max}$ of the magnetic field in the region where radio waves are created. Let us assume that radio waves are created at some radius $r_\gamma > r_0$. In this case $B_{\max} = B_0 (r_0/r_\gamma)^3$, while kink in the spectrum appear at energy
\begin{equation}
    \omega_{\text{kink}} \approx \sqrt{\frac{C_1 \langle|\cos\theta|\rangle}{C_2 B_{\max}}}, \qquad m_a^2 \ll \omega_p^2.
\end{equation}

\begin{figure}[t]
    \centering
    \includegraphics[width=0.5\linewidth]{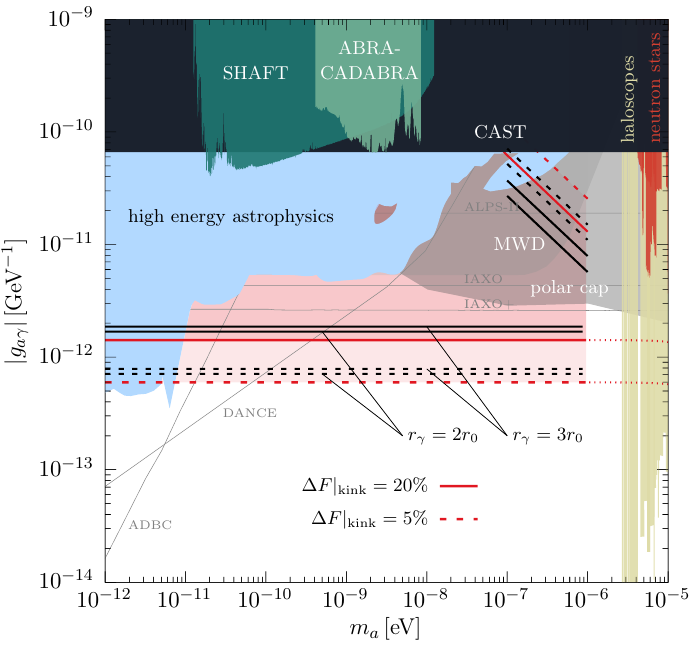}
    \caption{Dependence of sensitivity on the assumed location of radio-wave creation. In the main text it is assumed that it happens close to the surface. If the radio signal is instead created at a significant distance $r_\gamma = 2 r_0$ or $r_\gamma = 3 r_0$ from the NS surface (as labeled), the lower boundaries of the sensitivities change only very mildly. We also show an estimate for the change of the upper bound.    
    \label{fig:rgamma}}
\end{figure}

The bound of the sensitivity is defined by the condition that the probability of the resonant conversion is equal to some minimal value, $P_{\text{tot}} = P_{\min}$. This corresponds to the condition on the $P_{\text{lin}}$,  
\begin{equation}
    P_{\text{lin}} = - \frac{1}{3} \ln\left(1 - 3 P_{\min}\right).
\end{equation}
At the lower bound of sensitivity, the contribution of $B_+$ resonance dominates over $B_-$, so we can neglect the latter. The maximal linear probability is at $r = r_\gamma$ and it is given by
\begin{equation}
    P_{\text{lin}} = \frac{\pi g_{a\gamma}^2 \omega_{\text{kink}}}{m_a^2} \langle \sin^2 \theta\rangle B_{\max}^2 R,
\end{equation}
so the minimal coupling constant the one can probe is
\begin{equation}
    |g_{a\gamma}|_{\min} = \sqrt{\frac{P_{\text{lin}} m_a^2}{\pi \langle \sin^2 \theta\rangle \omega_{\text{kink}} B_{\max}^2 R}}.
\end{equation}
The dependence of the R-factor on radius is quite nontrivial. We remind that
\begin{equation}
    R = \left| \frac{d\ln m_{\text{eff}}^2 }{d r} \right|^{-1}_{r = r_\gamma}, \qquad m_{\text{eff}}^2 = \omega_p^2 - C_2 \omega^2 B^2.
\end{equation}
Using this definition and condition of the resonance $m_a^2 = m_{\text{eff}}^2$ one can obtain the following expression for R-factor,
\begin{equation}
    R = \frac{m_a^2}{\omega_p^2} \left| \frac{d\ln B }{d r} \right|^{-1}_{r = r_\gamma}, \qquad m_a^2 \ll \omega_p^2.
\end{equation}
Taking into account that $\omega_p^2 \propto B_{\max}$ we get
\begin{equation}
    |g_{a\gamma}|_{\min} \propto \frac{1}{B_{\max}^{1/4}} \left| \frac{d\ln B }{d r} \right|^{1/2}_{r = r_\gamma} \propto r^{1/4}_\gamma.
\end{equation}
We see that in our model the dependence of the lower bound on $r_{\gamma}$ is quite weak. Even if radio emission is created at $r_{\gamma} = 10 r_0$ the lower bound is relaxed only by a factor $10^{1/4}\approx 1.8$. To illustrate this point we show in Fig.~\ref{fig:rgamma} the change of the region of sensitivity for $r_\gamma = 2 r_0$ and $3 r_0$.

\section{Far side contribution to the signal}

\begin{figure}
    \centering
    \includegraphics[width=\textwidth]{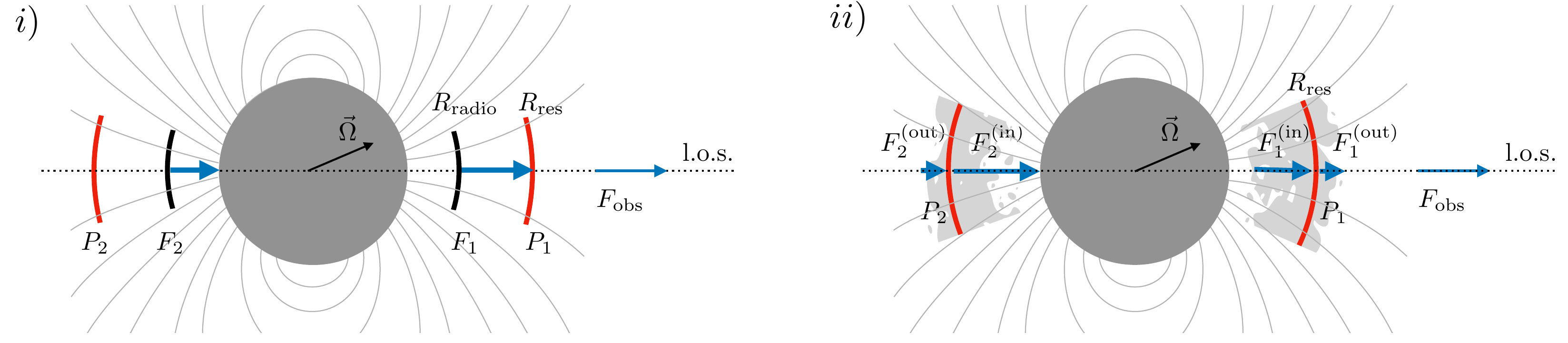}
    \caption{Two principal options for \textcolor{red}{the location of the resonance surface with respect to the surface of the radio photon creation. The left panel corresponds to our main assumption $R_{\text{radio}}< R_{\text{res}}$, while on the right panel radio signal is created in the extended gray area.}}
    \label{fig:radioscheme}
\end{figure}

Here we discuss the dependence of the signal prediction on the typical radius where the original radio signal is created, $R_{\rm radio}$ vs.~the location of the $B_+$ resonance, $R_{\rm res}$ for which we include a discussion on the potential contribution from the far side of the NS. We take a simplistic picture of spherical symmetry. This will make the arguments easily tractable and maximize canceling effects that are rooted in the geometry. There are various configurations possible, and in Fig.~\ref{fig:radioscheme} we highlight two principial ones. Left panel $i)$ corresponds to our principal assumption that $R_{\rm radio}< R_{\rm res}$. The right panel $ii)$ highlights a situation where the radio emission occurs in an extended zone indicated by the gray regions. Throughout we assume it to be a localized phenomenon in the sense that we are not required to perform integrals over extended angular regions.

In case $i)$ when $R_{\rm radio}< R_{\rm res}$ an emitted radio flux $F_1$ at radius $R_{\rm radio}$ becomes processed at $ R_{\rm res}$ above a threshold frequency with conversion probability $P_1$. The observed flux is then $F_{\rm obs}=F_1\left(1-P_1\right)$.
As a comparative measure to other cases, we may consider the flux decrement  at some characteristic frequency, i.e., by comparing the flux without conversion  $F_{\text {obs }}^{(P=0)}$ with the flux that would emanate when resonant conversion is possible, $F_{\text {obs }}^{(P\neq 0)}$,
\begin{align}
\Delta F \equiv F_{\text {obs }}^{(P\neq 0)}-F_{\text {obs }}^{(P=0)}=-P_1 F_1 .
\end{align}
This is the flux-decrement that we advertise as a signature in the main text. Under the current circumstances there is no contribution from the ``far side''.

The situation changes when we reverse the roles of both radii, and assume $R_{\rm radio} >  R_{\rm res}$. As is evident, no feature can be imprinted from the radio component that reaches us from the near side of the neutron star. However, denoting far side quantities by index ``2'', the observed flux is $ F_{\text {obs }}=F_1+F_2 P_2 P_1$ where the second term is from axions that traverse the NS before being regenerated as photons and a non-vanishing flux {\it increment} is present even in this case, 
\begin{align}
\Delta F = P_1 P_2 F_2 .
\end{align}

We may now turn to case $ii)$ with an extended zone of radio emission with the radius of resonance situated within this zone. Then we have two contributions to the unprocessed radio signal on each side. One generated within $r<R_{\rm res}$ denoted by $F^{\rm (in)} $ and one generated in the exterior $r>R_{\rm res} $ of the resonance zone,  $F^{\rm (out)} $. The observed flux then becomes
$ 
F_{\rm obs}=F_1^{\rm (out)}+F_1^{\rm (in)}\left[1-P_1\right]+F_2^{\text {\rm (out})} P_2 P_1
$. The change in photon flux then has contributions of both signs, 
\begin{align}
\Delta F  = -F_1^{\rm (in)} P_1 + F_2^{\text {\rm (out})} P_2 P_1.\end{align} The far side contribution may hence ``wash out'' the signal or, in a fine-tuned situation, cancel it altogether. 

Finally, we mention that we assumed a geometric ``obscuration limit'' for the far side. However, if $R_{\rm NS}\ll R_{\rm radio},\ R_{\rm res}$ both near and far side contribute on similar footing. This limit is of less interest for the current purposes as we rely on the resonance $B_+$ that is found in vicinity of the NS surface. Therefore, as stated in the main text, our proposal is sensitive to the assumption that $R_{\rm radio}$ is of similar order than $R_{\rm NS}$.

\end{document}